\documentclass{jfm}
\usepackage{graphicx}
\usepackage{subfigure,rotating,amsmath,float,upmath}
\usepackage{IEEEtrantools}
\usepackage{natbib}
\usepackage{color}
\ifCUPmtlplainloaded \else
  \checkfont{eurm10}
  \iffontfound
    \IfFileExists{upmath.sty}
      {\typeout{^^JFound AMS Euler Roman fonts on the system,
                   using the 'upmath' package.^^J}%
       \usepackage{upmath}}
      {\typeout{^^JFound AMS Euler Roman fonts on the system, but you
                   dont seem to have the}%
       \typeout{'upmath' package installed. JFM.cls can take advantage
                 of these fonts,^^Jif you use 'upmath' package.^^J}%
      }
  \else
  \fi
\fi

\ifCUPmtlplainloaded \else
  \checkfont{msam10}
  \iffontfound
    \IfFileExists{amssymb.sty}
      {\typeout{^^JFound AMS Symbol fonts on the system, using the
                'amssymb' package.^^J}%
       \usepackage{amssymb}%

      }{}
  \fi
\fi

\ifCUPmtlplainloaded \else
  \IfFileExists{amsbsy.sty}
    {\typeout{^^JFound the 'amsbsy' package on the system, using it.^^J}%
     \usepackage{amsbsy}}
    {\providecommand\boldsymbol[1]{\mbox{\boldmath $##1$}}}
\fi

{}

\def \Ca  {\mbox{Ca}}

\def \bn {\mathbf{n}}
\def \bu {\mathbf{u}}

\def\varepsilon{\epsilon}

\def \lp  {\left(}
\def \rp  {\right)}

\def \Uinf {U^{\infty}}
\def \Udp  {U_d}
\def \Ud   {U_{d}}
\def \Cadp {Ca_{d}}
\def \Cainf {Ca^{\infty}}
\def \rdh {R/H}
\def \hmin {h_{min}}
\def \hbar {\bar{h}}
\def \hthe {h^{EB}}
\def \hsim {h^{sim}|_{2D}}
\def \hsimmin {h_{min}^{sim}|_{2D}}

\title[A pancake droplet in a Hele-Shaw cell]{A pancake droplet translating in a Hele-Shaw cell: lubrication film and 
flow field}

\author
 {
 Lailai Zhu\aff{1}
  \corresp{\email{lailai.zhu@epfl.ch}},
  \and 
  Fran\c{c}ois Gallaire\aff{1}
  \corresp{\email{francois.gallaire@epfl.ch}},
  }
  \affiliation
{
\aff{1}
Laboratory of Fluid Mechanics and Instabilities, Ecole Polytechnique F\'ed\'erale de Lausanne, \\
Lausanne, CH-1015, Switzerland
}

\begin{document}
\maketitle 

\begin{abstract}
We adopt a boundary integral method to study the dynamics of a translating droplet confined in a Hele-Shaw cell in 
the Stokes regime. The droplet is driven by the motion of the ambient fluid with the same viscosity. We 
characterize the 
three-dimensional ($3$D) nature of the droplet interface and of the flow field.
The interface develops an arc-shaped ridge near the rear-half rim with a protrusion in the rear and a 
laterally symmetric pair of higher peaks; this pair of protrusions has been identified 
by recent experiments~\citep{huerre2015droplets} and predicted asymptotically 
~\citep{burgess1990analysis}. The  mean film thickness is well predicted by the extended 
Bretherton model~\citep{klaseboer2014extended} with fitting parameters. The flow
in the streamwise wall-normal middle plane is featured with recirculating zones,
which are partitioned by stagnation points closely resembling those of a two-dimensional droplet in a channel. 
Recirculation is absent in the wall-parallel, unconfined planes, in sharp contrast to the interior flow inside a 
moving  droplet in free space. The preferred orientation of the recirculation results from the anisotropic confinement 
of the Hele-Shaw cell. On these planes, we identify a dipolar disturbance flow field induced by the travelling 
droplet and its $1/r^2$ spatial decay is confirmed numerically. We pinpoint 
counter-rotating streamwise
vortex structures near the lateral interface of the droplet, further highlighting the complex $3$D flow pattern.

\end{abstract}

\section{Introduction}\label{sec:introduction}
The dynamics of a droplet or bubble pushed by a carrier fluid flowing in a confined space
is a classical multiphase problem that has a long history. In such cases, a capillary interface 
develops between the immiscible droplet/bubble and the carrier fluid that wets the wall. A thin film is formed 
between the interface and the wall, lubricating the droplet/bubble. Despite knowledge of the fundamental
picture of the thickness of the film, 
the shape of the menisci or the velocity of the suspended phase, and 
regardless of the steadfast efforts initiated in the 1960s by ~\citet{taylor1961deposition} 
and ~\citet{bretherton1961motion}, investigating a bubble confined in a tube as the first step, the dynamics of 
translating droplets/bubbles under confinement is not yet well understood.

The existing literature focuses mainly on a moving droplet/bubble confined in a capillary tube or between two 
closely spaced parallel plates (Hele-Shaw cell). In the former case, ~\citet{taylor1961deposition} performed 
experiments by blowing air into a tube filled with a viscous liquid where the air forms a round-ended cylindrical 
bubble. He measured the bubble velocity $\Udp$ compared with the mean 
velocity $\Uinf$ of the underlying flow, showing its excess velocity $m = \left(\Udp-U_{\infty}\right) /\Udp$ as a 
function of the capillary number $\Cadp = \mu \Udp/\gamma$, where $\mu$ denotes the dynamic viscosity of the liquid and 
$\gamma$ the surface tension; he also predicted the presence of stagnation points in the flow ahead of the front
meniscus and how the number and location of the stagnations vary with $m$. Almost at the same time, 
~\citet{bretherton1961motion} conducted similar experiments and performed an axisymmetric
lubrication analysis, showing that the lubrication equations were similar to their  
two-dimensional ($2$D) version assuming spanwise invariance. He focused on the shape of the front/rear menisci,  
the pressure drop, the thickness of the 
lubrication film and the excess velocity $m$. Bretherton established the well-known $2/3$ scaling between the 
non-dimensional film thickness $2h/H$ and the capillary number $\Cadp$, namely, $2h/H = P\lp 3\Cadp \rp^{2/3} $ with 
$P=0.643$, where $h$ and $H$ denotes the film thickness and the tube diameter respectively. 
The pre-factor $P$ could vary with the droplet/bubble's interfacial rigidity~\citep{bretherton1961motion, 
cantat2013liquid}, and the 
viscosity ratio between the droplet/bubble phase and the carrier phase~\citep{teletzke1988wetting}.

The situation is more complicated in a Hele-Shaw cell where the droplet is so squeezed that it adopts a flattened 
pancake-like shape, leaving a lubrication film between its interface and the wet plates (Fig.~\ref{fig:sketch}). 
Such flattened droplets are encountered in the context of droplet-based microfluidics~\citep{baroud2010dynamics} where 
droplets are manipulated 
in microfluidic chips to achieve micro-reaction, therapeutic agent delivery and biomolecule synthesis, 
etc~\citep{teh2008droplet}. Those chips are often thinner in the wall-normal direction than in others, in order to 
process simultaneously a large number of droplets constrained to move only horizontally. The problem of a moving 
pancake 
droplet in a Hele-Shaw cell hence serves as a model configuration to investigate the dynamics of those microfluidic 
droplets. 
Besides, the problem belongs to a larger set of research topics of moving menisci on a wet solid, a phenomenon that 
is involved in a broad range of industrial and natural situations~\citep{cantat2013liquid}
and has motivated 
 pioneering studies~\citep{park1984two,meiburg1989bubbles,burgess1990analysis} of the pancake 
droplet/bubble in a Hele-Shaw cell, as detailed below. 

The dynamics of the Hele-Shaw droplet/bubble  occur at different length scales spanning a 
broad range; their close coupling makes the problem truly multi-scale. The length scale in the unconfined direction is 
much larger than that in the confined direction. The latter corresponding
to the gap width of the cell is again much larger than the thickness of the lubrication film.
Thanks to the mathematical analogy between the governing equations of the depth-averaged Hele-Shaw flow and those
of the $2$D irrotational flow as proved by ~\citet{stokes1898mathematical} and commented by 
~\citet{lamb1932hydrodynamics}, potential flow theory was adopted  to study the motion of 
a 
Hele-Shaw bubble theoretically~\citep{taylor1959note} and numerically~\citep{tanveer1986effect}.
~\citet{park1984two} formulated a rigorous theory of a two-phase displacement problem (a less viscous fluid displacing 
a viscous one 
in a Hele-Shaw cell) as a double asymptotic expansion in small capillary numbers, $Ca$, and non-dimensional gap widths, 
$\epsilon$, of the cell (scaled by its transverse characteristic length scale); the theory 
holds as long as the viscosity ratio $\lambda$ between the displacing and displaced 
fluid satisfies $\lambda = o \lp Ca^{-1/3} \rp$.  
~\citet{burgess1990analysis} performed
a multi-region asymptotic analysis for a Hele-Shaw bubble based on the same assumption of small $Ca$ and $\epsilon$, 
focusing on the scaling dependence of the minimum/mean film thickness on  $Ca$ and $\epsilon$. Based on the stress jump 
derived by ~\citet{bretherton1961motion} and ~\citet{park1984two} that enables using lumped 
interfacial boundary conditions, depth-averaged $2$D simulations  
were carried out by 
~\citet{meiburg1989bubbles} for a Hele-Shaw bubble, including the leading-order effects of the dynamic 
meniscus hindering the movement of the bubble.
In a similar vein, an alternative depth-averaged framework has been 
recently implemented by ~\citet{nagel2015boundary} by solving the so-called $2$D Brinkman 
equations that take account of the in-plane velocity gradients.

These results are supposed to hold for a particular range of the parameter space due to their asymptotic nature 
and they have not been verified by either experiments or fully 
resolved $3$D
simulations.
Moreover, these studies often neglected the viscosity of the droplet phase or considered very low viscosities. The 
asymptotic 
analysis also fails to provide information such as the interior/exterior flow field, a full 
$3$D description of the droplet profile or lubrication film, or detailed connections with 
the droplet velocity. A tip of the iceberg has been revealed, and much effort will be 
required to reach 
a thorough understanding of the problem.
Very recently, elaborate experiments have been performed by ~\citet{huerre2015droplets} to measure the thickness and 
topology of the lubrication film between a viscous, surfactant-laden droplet and the wall. 
They identified a regime where the interface resembles a catamaran shape featuring two protrusions formed on 
its lateral sides, without providing a detailed explanation about its physical origin. Very few $3$D simulations 
have been conducted for a pancake
droplet/bubble despite the very recent work of ~\citet{ling15droplet} for a droplet with small but finite inertia. 
Here, we simulate a matching-viscosity droplet (the 
fluid inside and outside  has the same viscosity) in 
the inertialess regime based on an accelerated boundary integral method (BIM). We focus on the effect of 
the capillary number and the confinement (in other words the aspect ratio) of the droplet. We show the topology of the 
lubrication film and the spatial distribution of the film 
thickness. The dependence of the mean and minimum film thickness  on the capillary number are reported, and they 
are compared with the numerical and theoretical predictions of a $2$D droplet in a channel. Finally, 
we depict the flow field inside and outside the droplet, demonstrating its complex three-dimensionality.

\section{Problem description}\label{sec:problem}
\begin{figure}
\centering
	\includegraphics[scale=0.41]{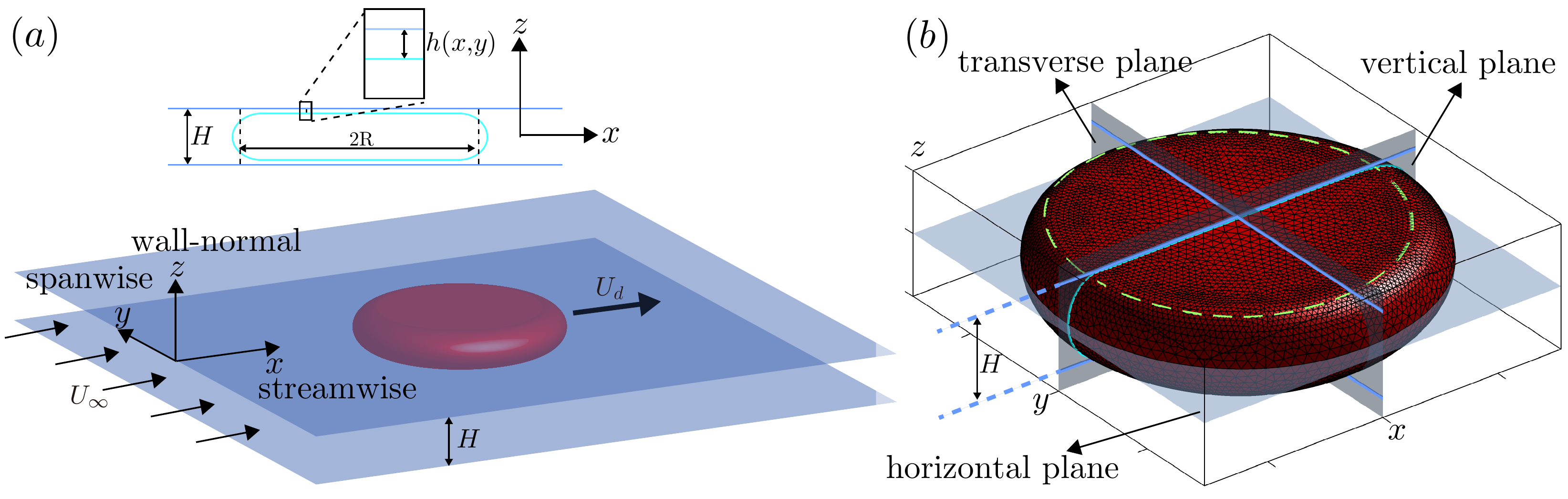}
		\caption{(a): A pancake droplet translating at velocity $\Ud$ in a Hele-Shaw cell with gap 
width $H$, driven by an ambient fluid with a mean velocity of $\Uinf$. The film thickness 
is $h(x,y)$ as denoted in the inset. (b): A discretized drop with $\Cainf=0.02$ under confinement 
$\rdh=2$. Blue lines denote the walls and the green dashed curve indicates the nearly-flat region of the film.}  
\label{fig:sketch}
\end{figure}

As shown in Fig.~\ref{fig:sketch} (a), we consider, in the creeping flow regime, a translating pancake droplet at 
velocity $\Ud$ driven by an ambient flow inside two infinitely large plates 
placed at $z=\pm H/2$. The fluids of the droplet phase and carrier phase are Newtonian, sharing the 
same dynamic viscosity $\mu$; the viscosity ratio $\lambda$ between the two (droplet phase versus carrier phase) is 
$1$. We solve the steady Stokes equations with no-slip boundary conditions on the plates and stress jump condition
$\boldsymbol{\sigma}_{1}\cdot \bn - \boldsymbol{\sigma}_{2}\cdot \bn =  \gamma \bn \lp 
\nabla_{S} \cdot \bn \rp$ on the droplet interface, where $\boldsymbol{\sigma}_{1}$ and $\boldsymbol{\sigma}_{2}$ 
are the total stress tensors corresponding to the carrier 
phase and drop phase respectively, $\bn$ is the unit normal vector on the interface pointing towards the carrier 
phase and $\nabla_{S}= \lp \mathbf{I} -\bn\bn \rp \cdot \nabla$ the surface gradient. A Poiseuille 
flow with a mean velocity of $\Uinf$ is applied in the inlet, hence the ambient velocity field in 
 is $\bu^{\infty} = \Uinf \left ( 1.5-6z^2/H^2,0,0\right)_{xyz}$. The radius 
of the droplet {at rest} is $a$ and all the length scales hereinafter are scaled by $a$ unless otherwise 
specified. Since 
the thickness $h(x,y)$ of the lubrication film is much smaller than the gap width $H$, the drop can be viewed as a 
cylinder of radius $R$ and height $H$, where 
$R^{2}H=4a^{3}/3$. We use $R/H$ to quantify the confinement. The surface 
tension of the droplet interface 
is $\gamma$. We define capillary numbers based on the velocity of the underlying flow or that of the 
droplet, leading to $\Cainf = \mu \Uinf/\gamma$ or $\Cadp = \mu \Udp/\gamma$ respectively. 

\section{Numerical methods}
We use a BIM accelerated by the general geometry Ewald method (GGEM) proposed by 
\citet{graham07_prl} and \citet{pranay2010pair}. On top of a GGEM-based BIM code originally developed to simulate 
elastic capsules in 
general geometries~\citep{zhu_sorting_14SM,lai14}, we implement a new module to simulate droplets.
Thanks to the linearity of Stokes equations, GGEM decomposes the flow field into two parts, a short-ranged, 
fast-decaying part solved
by traditional BIM techniques, and a long-ranged, smoothly varying part handled by a Eulerian
mesh-based solver for which we choose the spectral element method solver NEK5000~\citep{nek5000-web-page} here. 
For the 
details of our GGEM implementation, the reader is referred to ~\citet{lai14}. Our current 
work only accounts for a matching-viscosity droplet without the necessity for performing 
double-layer integrations, enabling us to follow directly the GGEM initially
developed for the fast computation of the Stokes flow driven by a set of point forces. To simulate a 
non-matching-viscosity droplet ($\lambda \neq 1$), we can further adopt the GGEM-accelerating BIM 
formulation~\citep{graham12_jcp} where the velocity field is expressed by  a single-layer integration solely even for 
problems with non-matching viscosities.

In the original GGEM-based BIM code for capsules, the interface is discretized by spherical 
harmonics. For the droplet interface, we use triangular elements instead for the discretization (see 
Fig.~\ref{fig:sketch} (b)). For a highly deforming interface that is far from a sphere, as in our case,
the triangular elements would capture the geometrical details more accurately and flexibly compared to the spherical 
harmonics. Another benefit of 
this choice is that adaptive mesh refinement on the interface like that performed in ~\citet{Zhu2013} can be readily 
incorporated to more efficiently and robustly describe the fine-scale geometrical features.

Based on the triangular elements, we perform singular integration on the droplet interface
using the plane polar coordinates with Gauss-Legendre quadrature, and a high-order 
near-singularity subtraction has also been adopted following~\citet{zinchenko2006boundary}. A robust fourth-order local
fitting algorithm (see Appendix B of ~\citet{zinchenko2006boundary} for details) is used to accurately calculate the 
surface normal vectors and curvatures of the interface. The most important feature
incorporated is the so-called passive mesh stabilization scheme~\citep{zinchenko2013emulsion} which has dramatically 
improved the robustness of our simulations because the orthogonality and 
smoothness of the triangular elements are well guaranteed over a long time evolution. For validation, we 
simulated a droplet tightly squeezed in a long tube and observed excellent agreement with the data of
~\cite{lac2009motion} based on a $3$D axisymmetric BIM implementation.

We used an open-source multiphase flow solver Gerris~\citep{popinet2009accurate} for some complementary 
simulations of a $2$D drop in a channel. Rigorous validations against our own $2$D BIM codes have been conducted.
Gerris is adopted here to obtain accurate flow fields conveniently.

\section{Results}\label{sec:results}
We focus on the regime $\Cainf\in \left(0.007, 
0.16\right)$ when the capillary forces are important. Lower capillary numbers are not pursued because they would 
require prohibitively high computational cost due to the rapid decrease of the 
film thickness $h$ with decreasing $\Cainf$. More precisely, numerical difficulties arise because of the 
singular perturbative nature of the problem at small $\Cainf$ values~\citep{park1984two}.
Three confinement levels $\rdh=1.5$, $2$ and $3$ have been examined; their corresponding gap widths 
are $H=0.840$, $0.693$ and $0.529$. As depicted in Fig.~\ref{fig:sketch}, we denote the $x$, $y$ and $z$ directions as 
the streamwise, spanwise and wall-normal directions, and the $yz$, $xz$ and $xy$ planes as the transverse, vertical and 
horizontal planes.

\subsection{Droplet velocity}\label{sec:udrop}
Fig.~\ref{fig:dropprofca002} (a) depicts the dependence of the scaled droplet velocity $\Udp/\Uinf$ with the 
capillary number $\Cainf$ and confinement $\rdh$. The velocity increases slightly with $\rdh$. This weak 
dependence is in accordance with the experimental observations
of ~\citet{shen2014dynamics} for  $\lambda \approx 1.4$ and capillary numbers several orders smaller than 
ours.
The scaled droplet velocity increases with $\Cainf$ monotonically and surpasses $1$, in contrast with the 
predicted velocity of 
$\Udp/\Uinf=1$ by ~\citet{gallaire2014marangoni} for a matching-viscosity pancake droplet modelled by an undeformed
cylinder at sufficiently low $\Cainf$. The mismatch results from two drawbacks of their model: it neglects the 
impeding effect of the dynamics menisci of the drop at low $\Cainf$; and it does not capture the film thickening 
at high $\Cainf$ that enhances the droplet velocity.

\subsection{Shape of the droplet and film thickness}\label{sec:shape}
\begin{figure}
    \hspace{0em}\includegraphics[scale = 0.475] {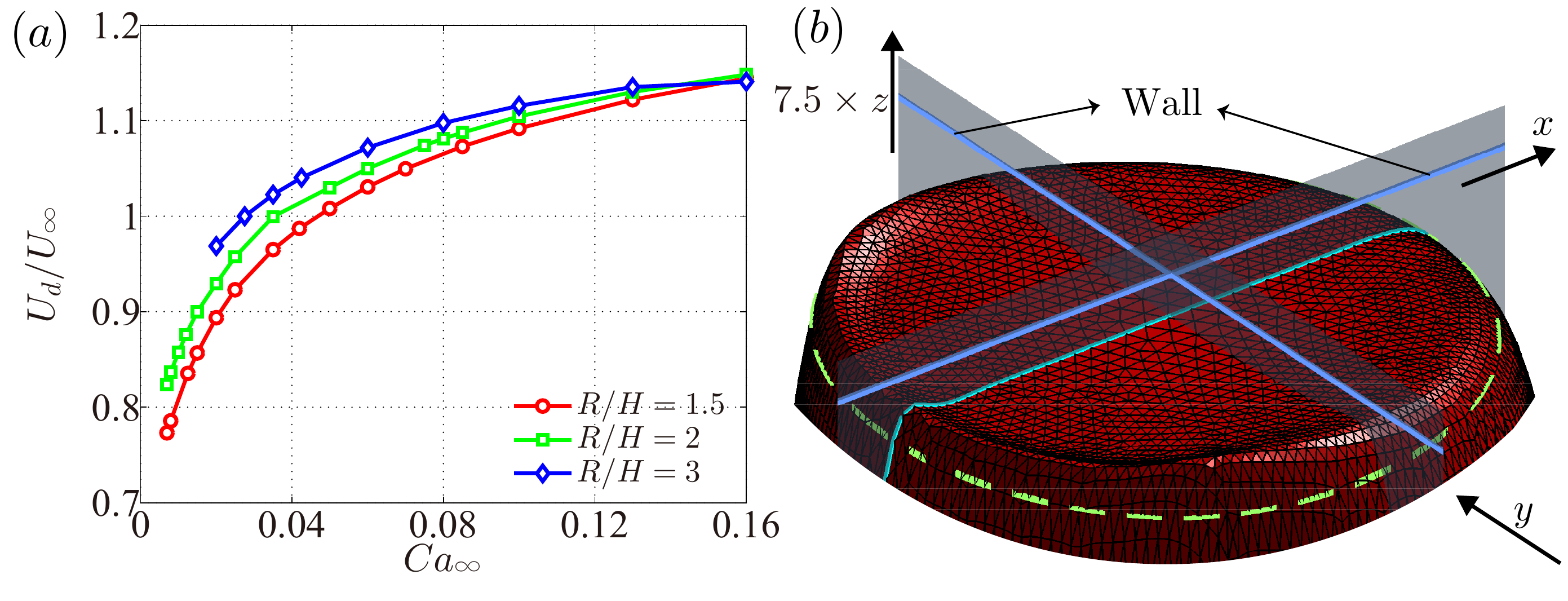}
       \caption{(a): The scaled droplet velocity $\Udp/\Uinf$ as a function of $\Cainf$ for varying confinement.
       (b): Stretching the thin film region of the drop as in Fig.~\ref{fig:sketch} (b) by $7.5$ times in $z$. }
\label{fig:dropprofca002}
\end{figure}

To better visualize the fine-scale geometrical features of the drop shown in Fig.~\ref{fig:sketch} (b), we stretch its 
top interface by $7.5$ times vertically and the zoomed view is shown in Fig.~\ref{fig:dropprofca002} (b).
The interface clearly bulges on the rear half of the rim of the interface, displaying an arc-shaped ridge.

\begin{figure}
    \hspace{-0.35em}\includegraphics[scale = 0.41] {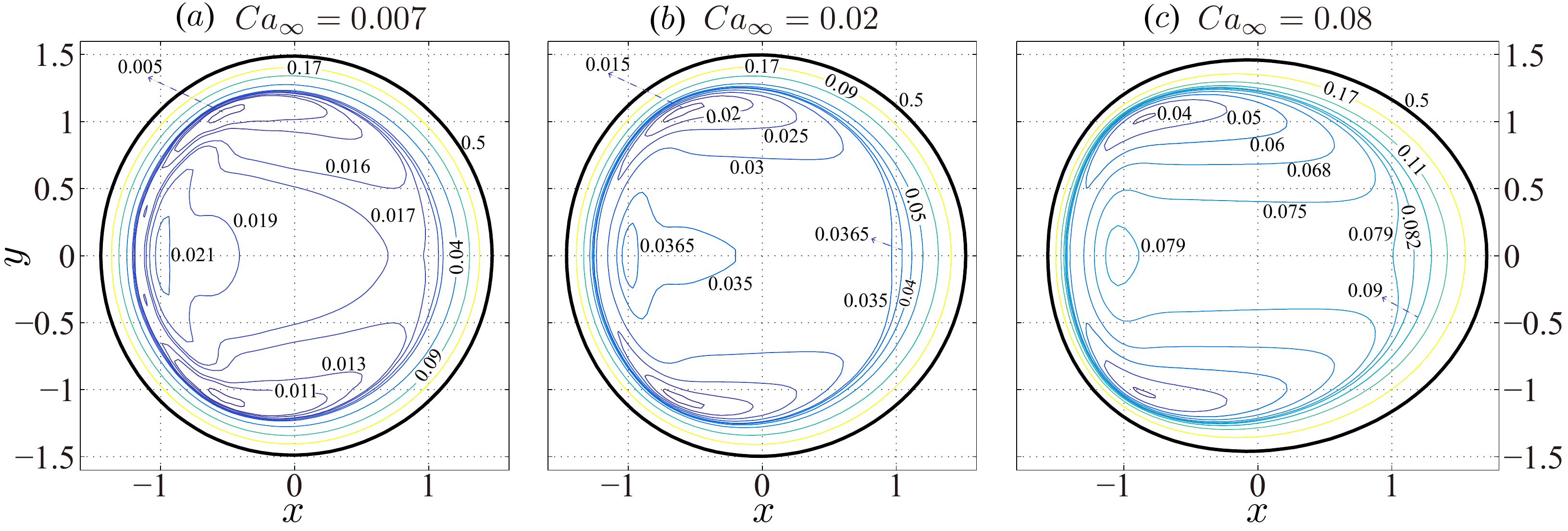}
       \caption{Contour lines of the scaled film thickness $h/H$ for droplets with $\Cainf=0.007$, $0.02$ and 
$0.08$  under confinement $R/H=2$. The black contour line $h/H=0.5$ indicates the edge of the droplet cut by the $z=0$ 
plane. }
\label{fig:filmthick}
\end{figure}

We show in Fig.~\ref{fig:filmthick} the contour lines of constant film thickness $h\lp x,y \rp/H$ for 
 droplets with $\Cainf=0.007$, $0.02$ and $0.08$ under confinement $\rdh=2$. 
Note that the height $z(x,y)$ of the droplet interface is inversely correlated to the film thickness $h(x,y)$, 
i.e. $z(x,y)+h(x,y)=H/2$.
The black curve $h/H=0.5$ represents the edge of the droplet cut by the $z=0$ plane, which resembles a circle at 
$\Cainf=0.007$ but becomes elongated at $\Cainf=0.08$.
For all $\Cainf$ investigated, the contour map exhibits three local minima: one at the rear and a 
symmetric pair on the lateral edges. These minima correspond to the peaks of the interfacial protrusions.
The two symmetric lateral protrusions are higher than the rear one. 
They have been recently observed experimentally  for a pancake droplet 
with $\lambda=25$ by ~\citet{huerre2015droplets}, who noted the resulting 'catamaran-like shape' adopted by the 
droplet. This feature has also been portrayed 
theoretically by ~\citet{burgess1990analysis}, performing a multi-region asymptotic analysis of a 
pancake bubble (see Fig. 5 of their paper). As far as we know, our study
represents the first computational work that identifies this unique interfacial topology.

~\citet{burgess1990analysis} showed in the low capillary number  limit that the contour lines of $h/H$ are 
streamwise parallel in the central film region (excluding the lateral portion) where the viscous 
forces dominate, resulting in the flat film. The contour lines of the $\Cainf=0.08$ case
are indeed parallel in the region $x \in (-1, 1), y\in(-0.75,0.75)$. At a reduced capillary number
$\Cainf=0.007$, such parallel lines disappear and the three protrusions instead
occupy a large portion of the film, pointing to its $3$D nature.

\begin{figure}
        \centering
 \hspace{-1em}\includegraphics[scale = 0.5] {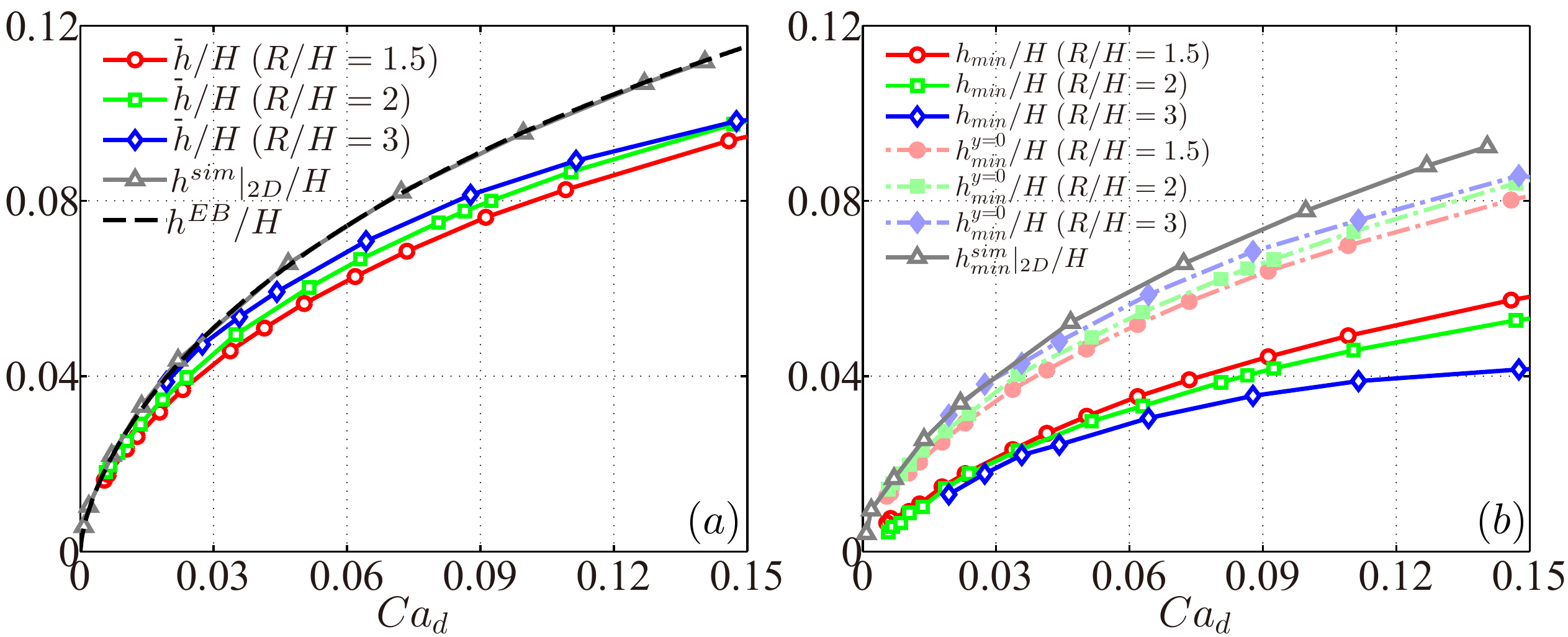}
       \caption{The scaled  mean $\hbar/H$ (a) and minimum $\hmin/H$ (b) film thickness versus
the capillary number $\Cadp$, for a pancake droplet under confinement $\rdh=1.5$ (circles), $2$ (squares) and $3$ 
(diamonds). Its minimum thickness on the middle vertical slice is denoted by 
$\hmin^{y=0}$. The dashed curve corresponds the constant film 
thickness $\hthe/H$ of a $2$D drop predicted by the EB model~\citep{klaseboer2014extended} with $P=0.6$ and $Q=1.5$.
The triangles denote the numerical data $h^{sim}|_{2D}/H$ (constant) and $h^{sim}_{min}|_{2D}/H$ (minimum) for a $2$D 
drop. 
} 
\label{fig:filmthick_overCa}
\end{figure}

We show in Fig.~\ref{fig:filmthick_overCa} (a) the dependence of the mean  thickness  $\hbar$ 
 on the capillary number.  $\Cadp$ is adopted instead of $\Cainf$ to be consistent with the prior studies. We obtain 
$\hbar$  by averaging $h$ over a central circular patch with radius 
$R_{cen}=0.3R_{xy}$, where $R_{xy}$ is the effective radius of the nearly circular droplet profile in the $z=0$ plane.
The scaled film thickness $\hbar/H$ increases with $\Cadp$ monotonically and weakly depends on $\rdh$.

For comparison, we use the flow solver Gerris to simulate a $2$D matching-viscosity droplet in a channel of 
width 
$H$ where the droplet length is much larger than its size in the confined direction. The film far away from the 
dynamic menisci is almost 
flat with a 
constant thickness of $\hsim$ which is reported in Fig.~\ref{fig:filmthick_overCa} (a). Additionally, we 
include the prediction of the extended Bretherton (EB) model proposed by ~\citet{klaseboer2014extended} for a
bubble, according to which, apart from the dynamic meniscus regions, the lubrication film has a constant thickness of 
$\hthe$
\begin{align}\label{eq:extend_bre}
 \hthe/H = \frac{1}{2}\frac{P \lp 3\Cadp \rp^{2/3}}{1+PQ\lp 3\Cadp \rp^{2/3}},
\end{align}
where $H$  is the tube diameter, and $P=0.643$ and $Q=2.79$ ~\citep{bretherton1961motion}.
This model agrees well with the empirical fit of ~\citet{aussillous2000quick} of 
~Taylor's (1961) experimental data. We adopt $P=0.6$ and $Q=1.5$ in Eq.~\ref{eq:extend_bre}, and 
the fitted thickness $\hthe/H$ almost coincides with the 
numerical value $\hsim/H$.
The mean film thickness $\hbar/H$ agrees well with the two values $\hsim/H$ and 
$\hthe/H$ of the $2$D drop at low capillary numbers, but starts deviating when $\Cadp$
increases. As the confinement increases, the film thickness $\hbar/H$ agrees better with the $2$D results.
The agreement between $\hbar/H$ with the thickness $\hsim/H \approx \hthe/H$ can be attributed to two reasons: first,
the central region where $\hbar$ is measured is rather flat as illustrated by the 
sparsely distributed contour lines in Fig.~\ref{fig:filmthick}, implying the mean film thickness 
$\hbar$ adopts the constant  thickness $h$ of the 
 vertical slice ($y=0$); second, as we will show in  section~\ref{sec:flowfield}, the velocity field 
of this slice strongly resembles that of a $2$D matching-viscosity droplet.

We plot in Fig.~\ref{fig:filmthick_overCa} (b) the scaled minimum film thickness $\hmin/H$ of the pancake droplet, 
where $\hmin^{y=0}/H$ denotes the scaled minimum thickness of its middle vertical slice, and 
 $\hsimmin/H$ that of the $2$D drop. For all 
$\rdh$, $\hmin^{y=0}/H$ is slightly below $\hsimmin/H$ and increases with $\rdh$. For the most 
confined case, $\rdh=3$,  $\hmin^{y=0}/H$ agrees with $\hsimmin/H$ reasonably well, which is in accordance
with the agreement between their mean thickness counterparts \textit{i.e.} $\hbar/H$ and $\hsim/H$ as discussed 
previously. 

The 
global minimum $\hmin/H$, is, however approximately half of the local $\hmin^{y=0}/H$, as 
can  be inferred from the minima of the contour maps (Fig.~\ref{fig:filmthick}) that 
represent the thickness of the film above the lateral and rear interfacial protrusions.
The difference between these two minima indicates the $3$D nature of the droplet interface. 
Note that, while $\hbar/H$ slightly increases with the confinement $\rdh$, $\hmin/H$ decreases significantly
with $\rdh$, especially at large $\Cadp$ numbers. This suggests that the $3$D nature is more pronounced for a more 
confined drop.

\subsection{Flow field in the reference frame of the droplet}\label{sec:flowfield}
\begin{figure}
    \hspace{-1em}\includegraphics[scale = 1.175] 
{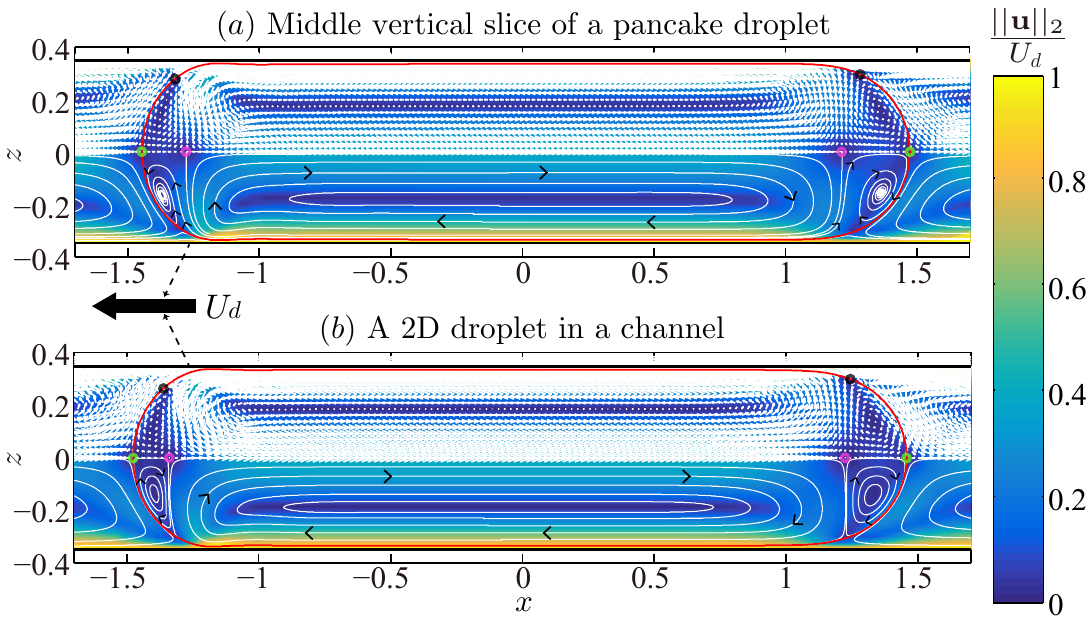}
       \caption{Velocity field in the droplet frame including the vectors and streamlines of 
the flow (a): on the $y=0$ plane of the drop with $\Cainf=0.007$ and $\rdh=2$.
        (b): of a $2$D droplet with $\Cainf=0.007$ and $\lambda=1$ travelling in a infinitely long channel. Red 
curves denote the droplet interface and black/magenta/tip circles denote the interfacial/axial/tip stagnation 
points; the contour colour indicates the in-plane velocity magnitude scaled by the droplet velocity 
$||\mathbf{u}||_{2}/\Udp$. }
\label{fig:vecxz}
\end{figure}

In this section, we focus on the flow field, $\bu_{drop} = \bu_{lab} - (\Ud,0,0)_{xyz}$, in 
the reference frame of the droplet, where $\bu_{lab}$ indicates that in the lab frame; the disturbance flow 
field will be discussed in section~\ref{sec:distfield}. The velocity fields projected on the vertical, 
horizontal and transverse 
planes in the reference frame of the drop are depicted. We first show in Fig.~\ref{fig:vecxz} (a) that on 
the middle vertical plane $y=0$ of the drop
with $\Ca=0.007$ under confinement $\rdh=2$. We compare it to the $2$D  
drop with $\lambda=1$ in Fig.~\ref{fig:vecxz} (b). We find the 
two flow patterns resemble each other closely, supporting the hypothesis made in section~\ref{sec:shape}
regarding their film thickness. In the top-half domain, the interior flow  consists of three recirculating zones, two 
clockwise ones appearing beside the front and rear meniscus respectively and a third anti-clockwise one in between; 
they are clearly distinguished by six stagnation points, two on the interface (black circles), two on the axis 
(magenta circles) and the other two as the tips (green circles) of the droplet. The front interfacial stagnation 
point has been predicted for an axisymmetric inviscid bubble in a tube by 
~\citet{taylor1961deposition}, as also discussed by~\citet{hodges2004motion}. The recirculation has been 
observed numerically by ~\citet{westborg1989creeping} and ~\citet{martinez1990axisymmetric} for an axisymmetric  
viscous droplet both near its front and rear meniscus, as well as by~\citet{ling15droplet} for a $2$D drop with 
$\lambda \approx 1.35$.

As explained by ~\citet{martinez1990axisymmetric}, this flow structure appears as a result of the combination of the 
shear exerted by the wall onto the film and the zero net flux condition inside the drop. The interface 
tends to follow the moving wall to reduce the viscous dissipation in the film, 
producing the interior backward flow; the zero net flux condition dictates a compensating 
forward flow in the near-axis region. This global balance results 
from the local divergence-free condition  $\partial u^{2D}_x /\partial x+\partial u^{2D}_z /\partial z=0$.

This $2$D scenario holds in any vertical slice of a spanwise, infinitely-long droplet confined by two plates.
But there is no reason why this condition should be satisfied in the middle slice
of the  `pancake'. The symmetry imposes indeed $u_y = 0$ but not necessarily $\partial u_y /\partial y = 0$. The 
similarity between the two flows shows \textit{a posteriori} that the in-plane divergence-free condition is 
approximately verified though, $\partial u_x /\partial x+\partial u_z /\partial z = -\partial 
u_y/ \partial y \approx 0$. This will be confirmed in the horizontal flow fields investigated next.

\begin{figure}
    \hspace{-0.75em}\includegraphics[scale = 0.47] {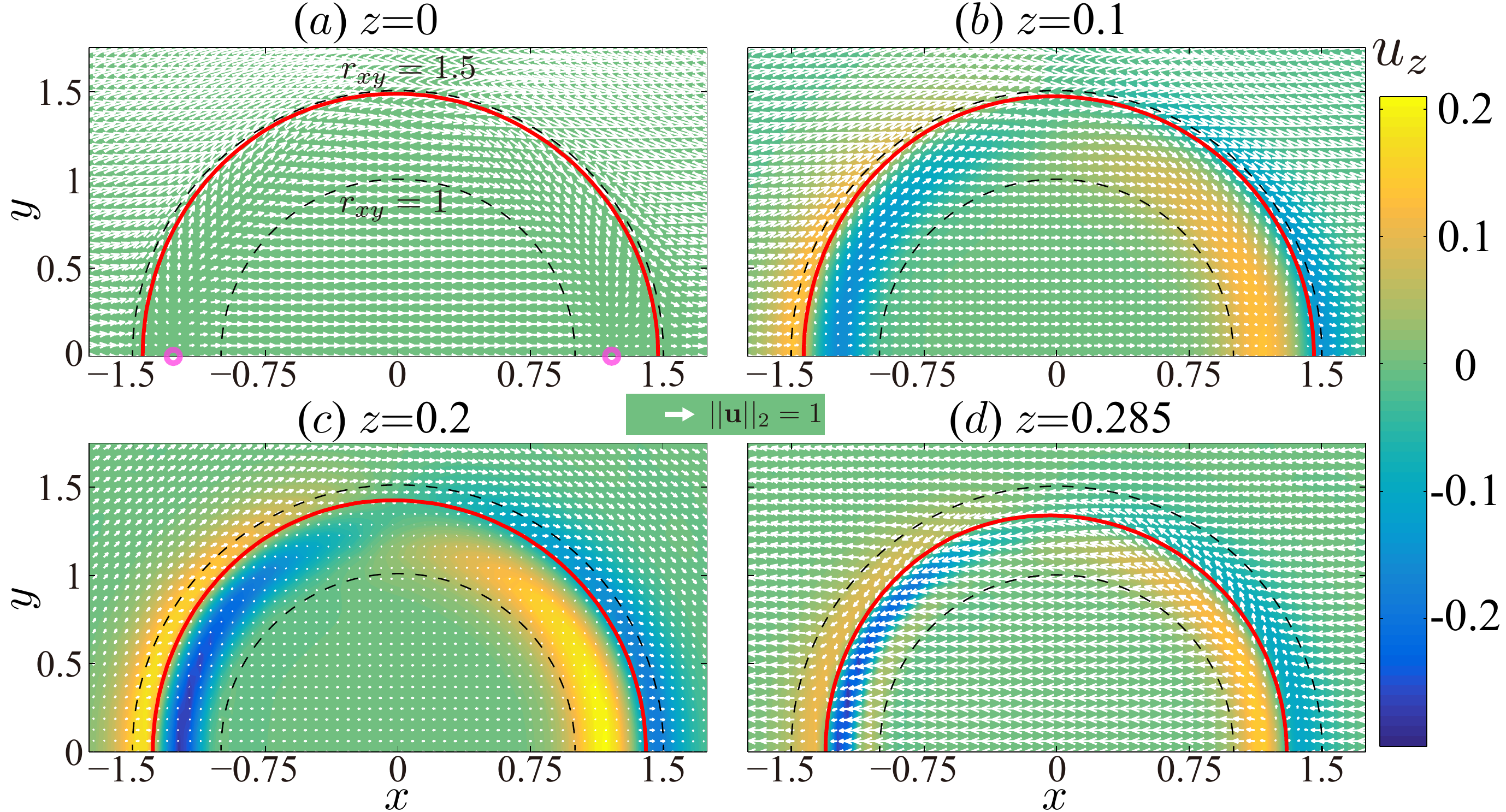}
       \caption{Flow on the horizontal planes at (a): $z=0$, (b): $z=0.1$, (c): $z=0.2$ and (d): 
$z=0.285$ for the same drop as in Fig.~\ref{fig:vecxz} (a), shown in 
half ($y \geqq 0$) of the domain. The top wall is located at $z=0.347$. The contour colour 
indicates the wall-normal velocity 
$u_{z}$. A reference vector with  norm $||\mathbf{u}||_2 = 1$ is given. Red curves
represent the droplet interface cut by the planes and the black dashed curves indicate the radial position of 
$r_{xy}=1$ and $r_{xy}=1.5$. Magenta circles in (a) denote the same axial stagnation points as in Fig.~\ref{fig:vecxz} 
(a).}
\label{fig:vecxy}
\end{figure}

In Fig.~\ref{fig:vecxy}, we display the velocity fields on the planes located at $z=0$, $0.1$, $0.2$ and 
$0.285$ together with the colour-coded wall-normal velocity $u_z$; note that the walls are located at $z =\pm 0.347$. 
The flow field can be partitioned into three patches depending on the 
radial position $r_{xy}$ with respect to the origin: first, the inner patch that is 
circular ($r_{xy} \lessapprox 1$) inside which the flow is mostly in the streamwise direction, \textit{i.e.}, $u_y 
\approx 0$ and $\partial u_y/\partial y \approx 0$; second, the outer patch ($r_{xy}\gtrapprox 
1.5$) that contains the flow passing around the droplet; and third, the annular patch ($1 \lessapprox r_{xy} 
\lessapprox 1.5$) that bridges the other two, where the flow mainly follows the in-plane curvature of the 
interface (red). 
The flow inside all the patches varies direction when the horizontal plane shifts from the 
middle $z=0$ towards
the top wall. More specifically, in the inner patch, the flow goes forward at $z=0$ but backward at 
$z=0.285$, reflecting the anti-clockwise recirculation on the vertical planes (see Fig.~\ref{fig:vecxz} (a)). In 
addition, the low in-plane velocities at $z=0.2$ correspond to the core of this 
recirculation. 
The velocity field in the outer patch represents the relative 
motion of the ambient flow with respect to the drop: near $z=0$, the flow is faster
than the drop and `pushes' it; near the wall, the flow is slower and `retards' it. 
The annular patch encompasses the droplet interface, and due to the non-penetration condition, the flow mostly follows 
the motion of the fluid elements along the interface: at $z=0$, the ambient flow `pushes' the droplet forward, 
resulting in a clockwise annular flow; 
near the top wall, the ambient flow `drags' the droplet backward resulting in a counter-clockwise flow. 
Unlike the middle vertical slice, the in-plane divergence-free condition in the middle horizontal plane is clearly
broken, as a source (resp. a sink) emerges on the axis at $x \approx -1.3$  (resp. $x \approx 1.2$) 
which exactly corresponds to the back (resp. the front) axial stagnation point on the middle vertical plane (see 
Fig.~\ref{fig:vecxz} (a)).

\begin{figure}
    \hspace{-1em}\includegraphics[scale = 0.345] {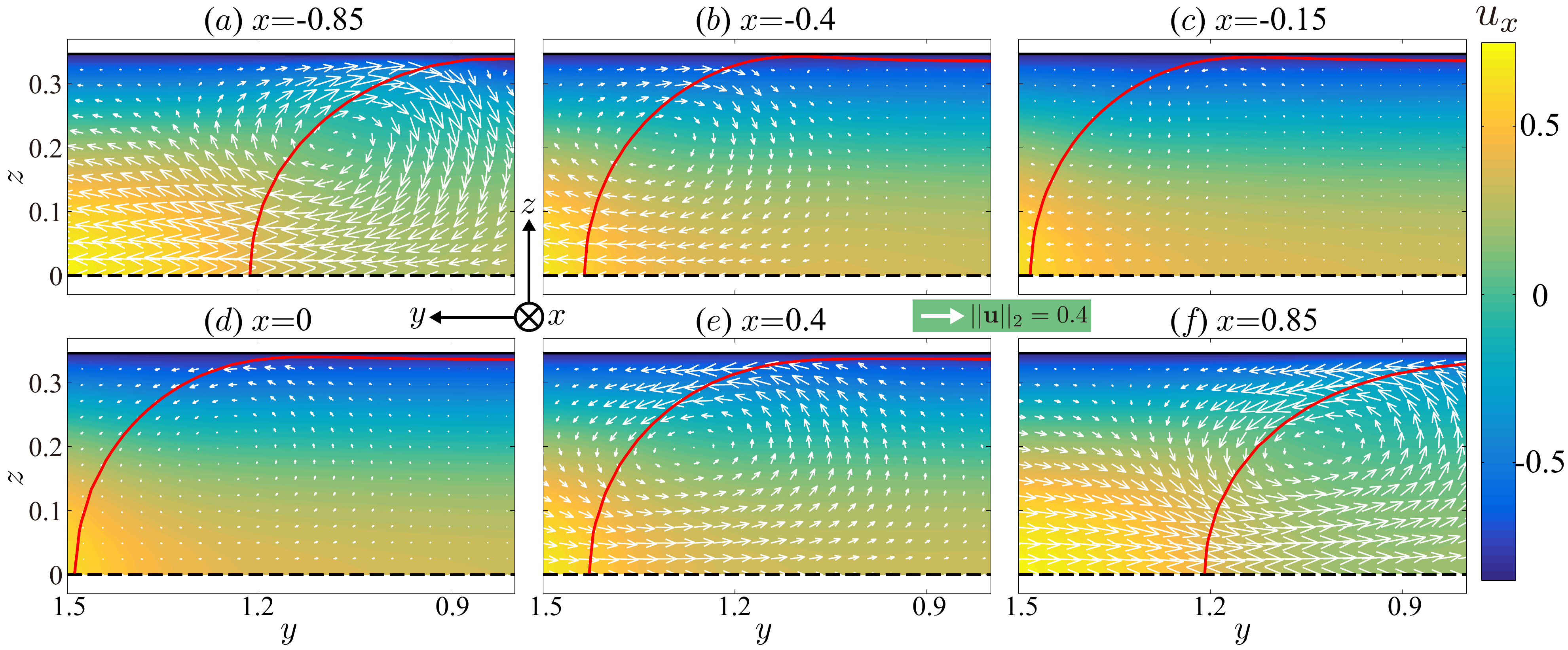}
       \caption{Flow on the transverse planes at (a): $x=-0.85$, (b): $x=-0.4$, (c): $x=-0.15$, 
(d): $x=0$, (e): $x=0.4$ and (f): $x=0.85$ for the same drop as in Fig.~\ref{fig:vecxz} (a), illustrated near 
the droplet interface (red) in the $y\geqq 0, z\geqq 0$ quarter of the domain. The contour 
 colour indicates the streamwise velocity $u_{x}$. A reference vector with  norm $||\mathbf{u}||_2 = 0.4$ is given.}
\label{fig:vecyz}
\end{figure}

We then come to the flow in the transverse planes shown in Fig.~\ref{fig:vecyz}. 
Because of  symmetry, we focus on the quarter ($y\geqq 0, z\geqq 0$) and we zoom in the 
lateral interface of the drop. We observe two vortical structures aligned in the streamwise direction:
one at the rear, rotating clockwise, and the other in the front, rotating anti-clockwise.
The two structures are most intense at approximately $x=-0.85$ and $0.85$, \textit{i.e.}, where their axis intersects 
the 
interface; they both decay in strength away from these maximum swirl regions and are connected at a no-swirl  position 
slightly aft the droplet centre, \textit{i.e.}, between the $x=-0.15$ and $x=0$ plane. At this position, the vorticity 
switches sign and streamlines change their spiralling direction. 
These streamwise vortex structures are closely related to the flow in the horizontal planes shown in 
Fig.~\ref{fig:vecxy}: at $x=-0.85$ and $y \approx 1$, the flow is in the 
positive (resp. negative) $y$ direction in the annular patch at $z=0$ (resp. $z=0.285$), which generates a 
clockwise vortex; the vortex at $x=0.85$ appears likewise though oppositely oriented, because the flows in the annular 
patch reverse their spanwise directions.

\subsection{Disturbance flow field}\label{sec:distfield}
\begin{figure}
    \hspace{0em}\includegraphics[scale = 1.05] {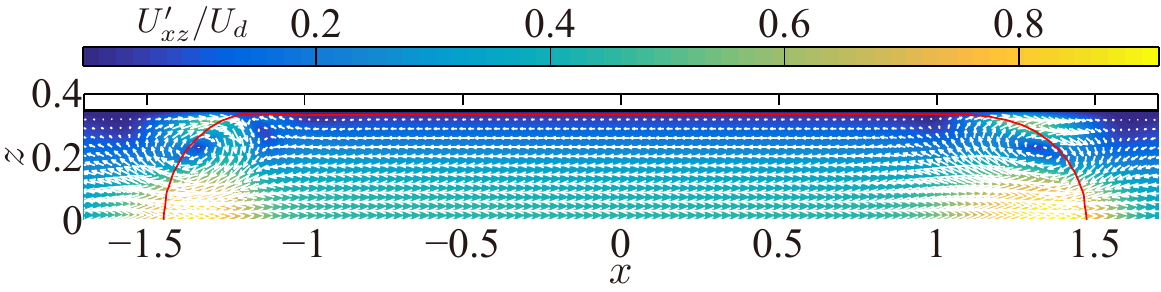}
       \caption{Disturbance flow field on the $y=0$ plane of the same droplet as that analysed in 
section \ref{sec:flowfield}.}
\label{fig:vecxz_dist}
\end{figure}

We hereby analyse the disturbance flow  $\bu^{\prime} = \bu_{lab} - \bu^{\infty}$ induced by the presence of a
translating pancake droplet, where $\bu^{\infty} = \Uinf \left ( 1.5-6z^2/H^2,0,0\right)_{xyz}$. For the same drop as 
that examined in section \ref{sec:flowfield}, we depict
$\bu^{\prime}$ on the middle vertical plane in Fig.~\ref{fig:vecxz_dist}. In most of the domain, the disturbance 
flow is parallel, in the direction against the underlying flow. This represents the obstructive effect of the droplet 
travelling at a velocity $\Ud$ smaller than the mean flow velocity $\Uinf$; in other words, the extra pressure drop
stemming from the presence of the droplet is positive. Interestingly, the disturbance flow $\bu^{\prime}$ reverses its 
 direction near the front and rear dynamic meniscus regions that extend from the lubrication film towards the 
static meniscus regions. 
As a result, two  vortical 
structures aligned in the positive $y$ direction emerge, akin to those observed in the 
flow field in the droplet 
frame $\bu_{drop}$ projected on the transverse ($yz$) planes as shown in Fig.~\ref{fig:vecyz}. In fact, the projections
of $\bu^{\prime}$, $\bu_{lab}$ and $\bu_{drop}$ on the transverse planes are equivalent,
because both the droplet velocity and the underlying flow
$\bu^{\infty}$ have only one non-zero component that is the $x$ component.

\begin{figure}
    \hspace{0em}\includegraphics[scale = 0.47] {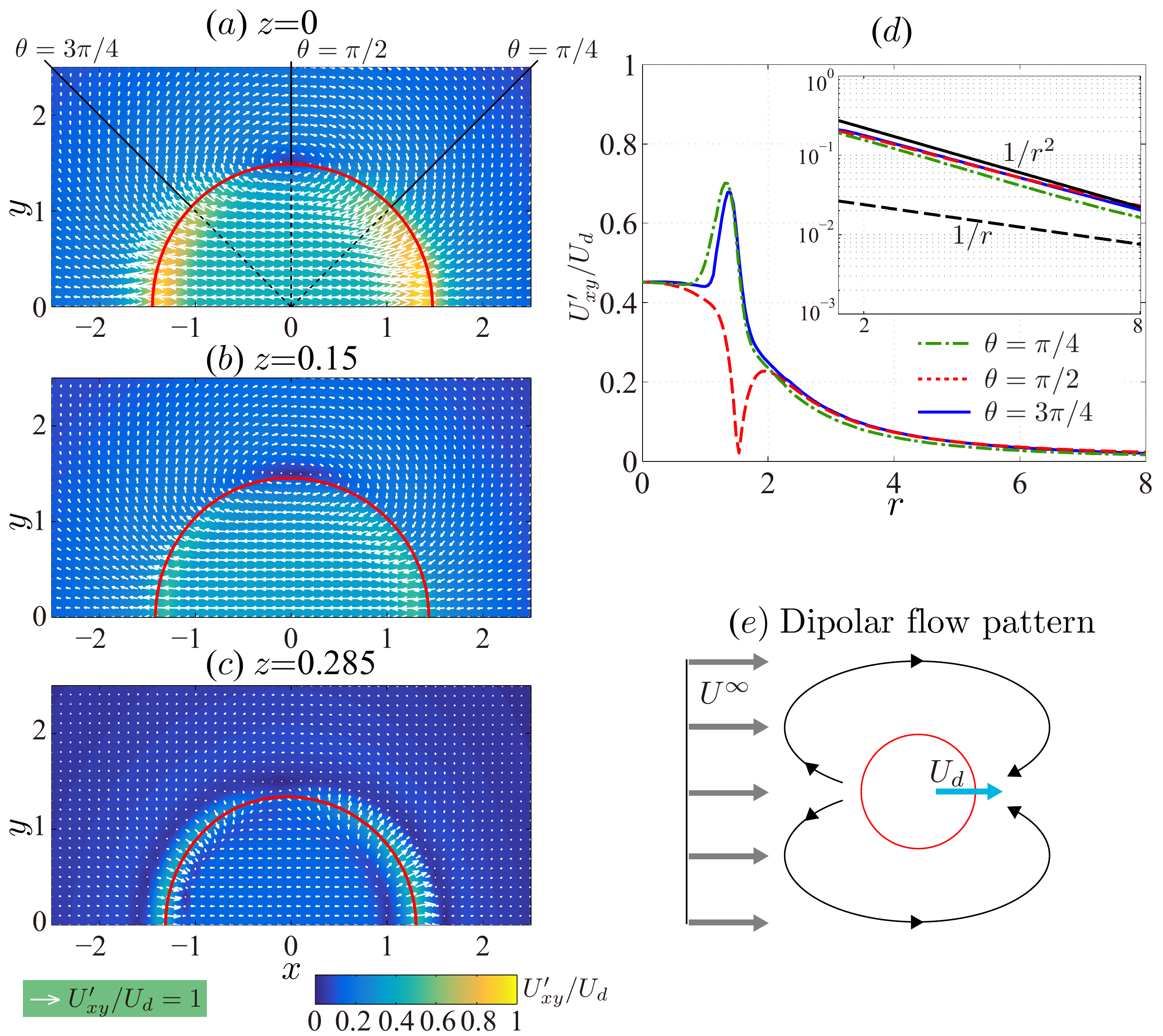}
       \caption{Disturbance flow field $\bu^{\prime}$ projected on the horizontal planes at 
(a): $z=0$, (b): $z=0.15$, (c): $z=0.285$ for the same drop as in Fig.~\ref{fig:vecxz_dist} (a); the contour colour
indicates the disturbance velocity magnitude $U^{\prime}_{xy}/\Ud$. (d): spatial variation of 
$U^{\prime}_{xy}/\Ud$ on the $z=0$ plane, 
along three directions; the inset shows the log--log scale.
(e): sketch of a typical dipolar flow pattern.}
\label{fig:vecxy_dist}
\end{figure}

The disturbance flow field $\bu^{\prime}$ projected on three horizontal planes is shown in
Fig.~\ref{fig:vecxy_dist}. On the middle $z=0$ plane, the droplet sucks in/ejects fluid in the front/rear,
the interior flow is mostly parallel and opposite to the moving direction of the droplet but reverses the sign near its
lateral edge. This resembles a $2$D dipolar flow field decaying as $1/r^2$ (see Fig.~\ref{fig:vecxy_dist}e for 
a typical sketch), which has been observed experimentally for a pancake droplet by ~\citet{beatus2006phonons}. This 
dipolar field, as an elementary solution of potential flow,  was also assumed to predict the velocity of a 
buoyancy-driven bubble ~\citep{maxworthy1986bubble}.
In Fig.~\ref{fig:vecxy_dist} (d), we examine how the disturbance 
velocity magnitude $U^{\prime}_{xy}=\sqrt{\lp u^{\prime}_x\rp^2 + \lp u^{\prime}_y\rp^2}$ varies with the radial 
distance $r=\sqrt{x^2+y^2}$,
along the three paths emitting from the centre of the domain; the angles between these paths and the positive $x$ 
direction
are  $\theta=\pi/4, \pi/2$
and $3\pi/4$. The log--log plot in the inset indicates that the decaying rate does indeed closely follows the $1/r^2$ 
scaling
law. The dipolar flow field is also detected on the $z=0.15$ plane with a decreased strength. However, it disappears
on the $z=0.285$ plane where the droplet ejects/sucks in fluid near its front/rear meniscus; this reversed disturbance 
flow has in fact been revealed on the middle vertical plane in Fig.~\ref{fig:vecxz_dist}.

\section{Conclusions and discussions}
We report a $3$D computation of a translating pancake droplet in a 
Hele-Shaw cell. The cell gap width is around $0.5 \sim 0.85$ the radius of a relaxed drop 
and the capillary number is in the range $\left[0.007,0.16\right]$.
In droplet-based microfluidic applications, the capillary numbers are smaller than our values by
an order of one to two~\citep{shen2014dynamics, huerre2015droplets} and the droplets are generally more confined.
Still, we believe our computational study has taken a first step towards handling these realistic situations by 
extending the previously explored parameter space.

Our simulations together with the recent experiments by ~\citet{huerre2015droplets} and the prior asymptotic 
analysis by ~\citet{burgess1990analysis} confirm a common and unique interfacial topology of a pancake
droplet/bubble, \textit{viz.} a pair of protrusions formed symmetrically on the lateral rim of the rear-half 
interface. The viscosity ratios of the three studies are $\lambda=1$, $25$ and $0$ respectively, suggesting that this
topology is rather insensitive to the viscosity ratio. As a complementary clue, the work of 
~\citet{lhuissier2013levitation} is worth noting. They investigated experimentally and 
theoretically the levitation of an oil drop ($\lambda \approx 2500$) on a moving wall mediated by the air film between 
them, observing a ridge of minimum film thickness on the downstream and lateral sides; although not 
explicitly mentioned, three closed iso-contour patterns were revealed indicating the interfacial protrusions 
(see their video~\citet{APSFGD14-0056}). 

The velocity field in the vertical planes closely resembles that of a $2$D droplet in a channel, while an analogous 
resemblance is missing in the horizontal planes. For a $2$D unconfined droplet or a $2$D Brinkman model 
of the drop~\citep{gallaire2014marangoni} where the confinement of Hele-Shaw cell is depth-averaged, the interior flow 
pattern in the drop frame, is featured with two symmetric counter-rotating recirculation regions to 
satisfy the zero  net flux condition; the drop's lateral interfaces recede due to the backward viscous forces from 
the exterior flow and consequently the flow near the symmetry axis advances to ensure global balance. 
For a $3$D pancake droplet, this feature is, however, absent in the horizontal planes. Recirculation therefore takes 
place
in a preferential direction, in the 
vertical planes in which the drop is confined but not in the horizontal unconfined planes.
This preference results from the anisotropy of the wall confinement as the viscous forces on the 
droplet interface in the vertical planes overwhelm those active in the horizontal planes. Indeed, the lubrication film 
bridging the wall and the interface is so thin that the viscous effects in the 
former case play a dominant role in the determination of the flow pattern.

Despite the $3$D feature of the flow, we have recovered that 
a moving pancake droplet induces a dipolar disturbance flow that can be described by a $2$D 
velocity potential $\phi^{\prime}$.
The dipole and the potential characterizing the disturbance are $\mathbf{d} = \lp R^2\lp  \Udp 
-\Uinf\rp,0\rp_{xy}$ and $\phi^{\prime} = -\mathbf{d} \cdot \mathbf{r}/r^2$ respectively, where 
$\mathbf{r}$ is the position vector with respect to the droplet centre. This shows that the leading contribution of the 
disturbance flow, $\nabla \phi^{\prime}$, decays as $1/r^{2}$. This scaling is attributed to the confining effect of 
the two parallel walls and is important to 
bear in mind when considering the hydrodynamic interactions among several pancake droplets or among the droplets and 
the lateral boundaries in micro-fluidic chips.

Planned future work includes the analysis of force balance on the droplet determining its velocity 
based on the obtained $3$D data, as well as the extension of our 
GGEM-based BIM code to account for non-matching-viscosity droplets and interfacial transport of insoluble 
surfactants.

\section*{Acknowledgements}
We thank Dr. Etienne Lac for sharing the data of ~\citet{lac2009motion}. 
Dr. Mathias Nagel and Giacomo Gallino are acknowledged for performing $2$D BIM computations in 
support of validating our Gerris set-up. We thank Gioele Balestra for delightful discussions. This work was 
supported by a grant from the Swiss National Supercomputing Centre (CSCS) under project ID s603.
The European Research Council is acknowledged for funding the work through a starting grant (ERC SimCoMiCs 280117).

\bibliographystyle{jfm}

\end{document}